\documentstyle[12pt]{article}

\def\be{\begin{equation}}
\def\ee{\end{equation}}
\def\bea{\begin{eqnarray}}
\def\eea{\end{eqnarray}}

\input{tcilatex}

\begin{document}

\begin{titlepage}

\title{Discrete transformation for matrix 3-waves problem in three dimensional
space}
\author{A.N. Leznov$^{1,2,3}$ and R.Torres-Cordoba$^{1}$\\
$^1${\it Universidad Autonoma del Esatado de Morelos, CCICap}\\
$^2${\it  Institute for High Energy Physics, 142284 Protvino,}\\
{\it Moscow Region,Russia}\\
{\it and}\\
$^3${\it  Bogoliubov Laboratory of Theoretical Physics, JINR,}\\
{\it 141980 Dubna, Moscow Region, Russia}}
\maketitle

\begin{abstract}
Discrete transformation for 3- waves problem is constructed
in explicit form.  Generalization of this system on the matrix case in three
dimensional space  together with corresponding discrete transformation is
presented also.
\end{abstract}

\end{titlepage}

\section{Introduction}

The problem of three waves in two dimensions arises in different form in many
branches of the mathematical physics. Its application to problems of
radiophysics and nonlinear optics reader can find in \cite{1}. In connection
with the inverse scattering method it was investigated in \cite{2} and
considered in details in numerous further papers.

The goal of the present paper is generalize this system on the space of
three dimension with simultaneous exchanging the unknown scalar functions on
the operator valued ones. The last generalization allows to include into
consideration the quantum region with the Heisenberg operators as unknown
functions of the problem.

It is necessary to mention that numerous different (by the form) discrete
transformation were used up to now with respect to two components
hierarchies of integrable systems, which are connected with so called
Darboux-Toda, Lotke-Volterra and Heisenberg substitutions. In the present
paper we come to the substitution which connect six independent functions,
which corresponds to $A_{2}$ algebra but not to $A_{1}$ as it was in the
case of the two components systems. This substitution may be considered as
the integrable mapping, connected six initial function with the six final
ones. Substitutions of the present paper do not coincide with recently
introduced in \cite{3} so-called Ultra-Toda mappings. And the last comment.

The method of this paper without any difficulties can be generalized on the
case of n-th wave problem. In this case the number of independent variables
of substitution will be $(n\times (n+1)$ which coincides with the number of
positive and negative roots of $A_{n}$ algebra.

The traditional way for obtaining the system of equations for 3-waves
problem in $(1+1)$ dimensions is the $L-A$ pair formalism 
\[
[\partial_x-u,\partial_t-v]=0 
\]
with 
\[
u=%
\pmatrix{ c_1\lambda & (c_2-c_1)P & (c_3-c_1)Q \cr
                     (c_1-c_2)B & c_2\lambda & (c_3-c_2)A \cr
                     (c_1-c_3)D & (c_2-c_3)E & -(c_1+c_2)\lambda \cr}, 
\]
\[
v=%
\pmatrix{ d_1\lambda & (d_2-d_1)P & (d_3-d_1)Q \cr
                     (d_1-d_2)B & d_2\lambda & (d_3-d_2)A \cr
                     (d_1-d_3)D & (d_2-d_3)E & -(d_1+d_2)\lambda \cr} 
\]
where $c,d$ four arbitrary numerical parameters $(\sum c_i=0, \sum d_i=0)$
and $(A,B,D,E,P,Q)$ are unknown functions of the problem. The system of
equations for them has the form 
\[
(d_2-d_1)P_x-(c_2-c_1)P_t+\nu EQ=0,\quad (d_3-d_1)Q_x-(c_3-c_1)Q_t-\nu PA=0, 
\]
\begin{equation}
(d_3-d_2)A_x-(c_3-c_2)A_t+\nu QB=0,\quad (d_1-d_2)B_x-(c_1-c_2)B_t-\nu AD=0,
\label{3W}
\end{equation}
\[
(d_1-d_3)D_x-(c_1-c_3)D_t+\nu BE=0,\quad (d_2-d_3)E_x-(c_2-c_3)E_t-\nu DP=0, 
\]
where $\nu=3(c_1d_2-c_2d_1)$. Soliton like solution for the last system of
equations may be easily found with the help of the technique of the paper %
\cite{4}. From these results it can be found the first steps of the discrete
transformation with respect to which the last system is invariant. In
present paper the reader have to consider the form of the discrete
transformation as the lucky guess.

\section{Discrete transformation}

We can assume (but this is a direct corollary of the results of \cite{4})
that the three ''new'' functions $(Q,A,P)$, denoted by the bar symbols,
connected with the old ones as following 
\[
\bar{Q}={\frac{1}{D}},\quad \bar{A}=-{\frac{B}{D}},\quad \bar{P}={\frac{E}{D}%
} 
\]%
satisfy the system (\ref{3W}). Then from the first and second equations of
the first column it is possible to determine $\bar{E},\bar{B}$ functions
with the result 
\[
\bar{E}=-{\frac{1}{(c_{1}-c_{3})}}{\frac{E}{D}}(D_{x}-(c_{2}-c_{1})BE)+{%
\frac{1}{(c_{2}-c_{3})}}(E_{x}+(c_{2}-c_{1})DP), 
\]%
\[
\bar{B}={\frac{1}{(c_{1}-c_{3})}}{\frac{B}{D}}(D_{x}-(c_{2}-c_{1})BE)-{\frac{%
1}{(c_{1}-c_{2})}}(B_{x}-(c_{3}-c_{2})AD). 
\]%
And in a self-consistent way determine from the second and first equations
of the second column $\bar{D}$. We will not present here this not very
simple expression, because in few lines below we will have observable
expression for this value. Straightforward but tedious calculations show
that the third equation of the first column is also satisfied ($\bar{D}_{3}=-%
\bar{B}\bar{E}$).

For further consideration it is more suitable to introduce three dependent
variables $(\quad \xi+\eta+\sigma=0)$ 
\[
\xi=(d_2-d_1)t+(c_2-c_1)x,\quad \eta=(d_3-d_2)t+(c_3-c_2)x,\quad
\sigma=(d_1-d_3)t+(c_1-c_3)x. 
\]
In each pairs of variables $(\xi,\eta),(\xi, \sigma), (\eta,\sigma)$ the
differentiation operators take the form 
\[
\pmatrix{
\partial_1\equiv {(d_2-d_1)\over \nu}\partial_x-{(c_2-c_1)\over \nu}\partial_t \cr
\partial_2\equiv {(d_3-d_2)\over \nu}\partial_x-{(c_3-c_2)\over \nu}\partial_t \cr
\partial_3\equiv {(d_1-d_3)\over \nu}\partial_x-{(c_1-c_3)\over \nu}\partial_t
\cr}= 
\pmatrix{
-\partial_{\eta} & \partial_{\sigma} & \partial_{\sigma}-\partial_{\eta} \cr
\partial_{\xi} & \partial_{\xi}-\partial_{\sigma} &-\partial_{\sigma} \cr
\partial_{\eta}-\partial_{\sigma} & -\partial_{\xi}& \partial_{\eta} \cr} 
\]
Really the explicit form the generators of differentiation via $%
(\xi,\eta,\sigma)$ variables will be not essential. Now the system (\ref{3W}%
) looks much more attractive 
\[
P_1=-QE,\quad A_2=-BQ,\quad Q_3=-PA 
\]
\begin{equation}
B_1=-AD,\quad E_2=-DP,\quad D_3=-EB  \label{MS}
\end{equation}

In the last form the system is obviously invariant with respect to
permutation of the indexes of differentiation with the simultaneous
corresponding exchanging of unknown functions. The discrete transformation
of the beginning of this section may be rewritten in more symmetrical form
(we will denote it with the help of the symbol $T_{3}$) 
\[
\bar{Q}={\frac{1}{D}},\quad \bar{A}=-{\frac{B}{D}},\quad \bar{P}={\frac{E}{D}%
}, 
\]%
\[
\bar{B}=D({\frac{B}{D}})_{2},\quad \bar{E}=-D({\frac{E}{D}})_{1},\quad {%
\frac{\bar{D}}{D}}=DQ-(\ln D)_{1,2} 
\]%
By the permutation indexes $(1,3)$ (together with corresponding exchanging
of unknown functions) it is possible to obtain the $T_{1}$ discrete
transformation with respect to which the system (\ref{MS}) is also invariant 
\[
\bar{P}={\frac{1}{B}},\quad \bar{Q}={\frac{A}{B}},\quad \bar{E}=-{\frac{D}{B}%
}, 
\]%
\[
\bar{D}=B({\frac{D}{B}})_{2},\quad \bar{A}=-B({\frac{A}{B}})_{3},\quad {%
\frac{\bar{B}}{B}}=BP-(\ln B)_{2,3} 
\]%
And at last the discrete transformation $T_{2}$ has the form 
\[
\bar{A}={\frac{1}{E}},\quad \bar{B}={\frac{D}{E}},\quad \bar{Q}=-{\frac{P}{E}%
}, 
\]%
\[
\bar{D}=-E({\frac{D}{E}})_{1},\quad \bar{P}=E({\frac{P}{E}})_{3},\quad {%
\frac{\bar{E}}{E}}=EA-(\ln E)_{1,3} 
\]

In the form presented above substitutions $T_{i}$ may be considered as a
mapping, connected six initial (unbar) functions with six final (bar) ones.
From the other side each substitution may be considered as the infinite
dimensional chain of equations. For instance the corresponding chain of
equation in the case of $T_{1}$ substitution has the form 
\begin{equation}
{\frac{B^{n+1}}{B^{n}}}-{\frac{B^{n}}{B^{n-1}}}=-(\ln B^{n})_{2,3},\quad
D^{n+1}=B^{n}({\frac{D^{n}}{B^{n}}})_{2},\quad A^{n+1}=-B^{n}({\frac{A^{n}}{%
B^{n}}})_{3}  \label{CH}
\end{equation}%
\[
E^{n+1}=-{\frac{D^{n}}{B^{n}}},\quad Q^{n+1}={\frac{A^{n}}{B^{n}}} 
\]%
In the first row we have the lattice like system connected 3 unknown
functions $(B,D,A)$ in each point of the lattice. The first chain for $B$
functions is exactly well known two dimensional Toda lattice.

\section{Some properties of the discrete transformations}

All constructed above discrete transformations are invertible. This means
that unbar unknown function may presented in terms of the bar ones. For
instance $T_{3}^{-1}$ looks as 
\[
D={\frac{1}{\bar{Q}}},\quad B=-{\frac{\bar{A}}{\bar{Q}}},\quad E={\frac{\bar{%
P}}{\bar{Q}}}, 
\]%
\[
P=-\bar{Q}({\frac{\bar{P}}{\bar{Q}}})_{2},\quad A=\bar{Q}({\frac{\bar{A}}{%
\bar{Q}}})_{1},\quad {\frac{Q}{\bar{Q}}}=\bar{D}\bar{Q}-(\ln \bar{Q})_{1,2} 
\]

It is not difficult to check by direct computation that discrete
transformations $T_i$ are mutual commutative $(T_iT_j=T_jT_i)$ on the
solutions of the system (\ref{MS}).

We present below corresponding calculations to prove that $T_1T_2=T_2T_1=
T_3 $. Indeed result of the action of $T_1$ on some solution of the system (%
\ref{MS}) is the following 
\[
P^1={\frac{1}{B}},\quad Q^1={\frac{A}{B}},\quad E^1=-{\frac{D}{B}}, 
\]
\[
D^1=B({\frac{D}{B}})_2,\quad A^1=-B({\frac{A}{B}})_3,\quad {\frac{B^1}{B}}%
=BP-(\ln B)_{2,3} 
\]
Action of the $T^2$ transformation on this solution leads to 
\[
A^{21}={\frac{1}{E^1}}=-{\frac{B}{D}},\quad B^{21}={\frac{D^1}{E^1}}=D({%
\frac{B}{D}}), 
\]
\[
Q^{21}=-{\frac{P^1}{E^1}}={\frac{1}{D}},\quad D^{21}=-E^1({\frac{D^1}{E^1}}%
)_1= -{\frac{D}{B}}(B(\ln D)_2-B_2)_1=QD^2-D(\ln D)_{12} 
\]
\[
P^{21}=E^1({\frac{P^1}{E^1}})_3={\frac{E}{D}},\quad E^{21}=(E^1)^2A^1-E^1
(\ln E^1)_{13}=-D({\frac{E}{D}})_1 
\]
The same calculation repeated in the back direction shows that $W^{1,2}=
W^{2,1}=W^3$ - the result of application of the $T_3$ transformation to an
initial solution $W$.

Thus from each given initial solution $W_0\equiv (A,P,Q,E,B,D)$ of the
system (\ref{MS}) it is possible to obtain the chain of solutions labeled by
two natural numbers $(l_1,l_2$, or $(l_3))$ the number of application of the
discrete transformations $(T_1,T_2,T_3)$ to it (as it was shown above $%
T_1T_2= T_2T_1=T_3$).

The arising chain of equations with respect to $(D,B,E)$ functions are
exactly two dimensional Toda lattices. Their general solutions in the case
of two fixed ends are well-known \cite{LS}. As reader will be seen soon this
fact allows to construct the many soliton solutions of the 3-wave problem in
the most straightforward way.

\section{Resolving of discrete transformation chains}

\subsection{Two identities of Jacobi}

We begin from the following obvious equalities for determinants of n-th
order 
\[
Det_{n}(T_{n})\equiv D_{n}%
\pmatrix{ T_{n-1} & a \cr
                               b  & \tau \cr}=D_{n-1}(T_{n-1})(\tau
-bT_{n-1}^{-1}a)\equiv D_{n-1}(T_{n-1})\tilde{\tau} 
\]%
where $T_{n-1}$ is $(n-1)\times (n-1)$ matrix, $a,b$ are $(n-1)$ dimensional
column (row) vectors respectively and $\tau $ scalar.

By the same reason the following formula takes place 
\[
D_{n}%
\pmatrix{ T_{n-2} & a^1 & a^2 \cr
              b^1  & \tau_{11} & \tau_{12} \cr
              b^2  & \tau_{21} & \tau_{22} \cr}=D_{n-2}(T_{n-2})D_{2}%
\pmatrix{
\tau_{11}-b^1T^{-1}_{n-2}a^1 & \tau_{12}-b^1T^{-1}_{n-2}a^2 \cr
\tau_{21}-b^2T^{-1}_{n-2}a^1 & \tau_{11}-b^2T^{-1}_{n-2}a^2 \cr} 
\]%
where $a^{i},b^{i}$ are $(n-2)$ dimensional columns (rows) vectors, $\tau
_{i,j}$ components of 2-th dimensional matrix. It is obvious how relations
of these types may be continued.

Now using results above let us transform the following expression 
\[
D_{n}\pmatrix{ T_{n-1} & a^1 \cr
              b^1  & \tau_{11} \cr}D_{n}%
\pmatrix{ T_{n-1} & a^2 \cr
                                         b^2  & \tau_{22} \cr}-D_{n}%
\pmatrix{ T_{n-1} & a^2 \cr
              b^1  & \tau_{12} \cr}D_{n}%
\pmatrix{ T_{n-1} & a^1 \cr
                                         b^2  & \tau_{21} \cr}= 
\]%
\[
D_{n-1}^{2}(T_{n-1})D_{2}%
\pmatrix{
\tau_{11}-b^1T^{-1}_{n-1}a^1 & \tau_{12}-b^1T^{-1}_{n-1}a^2 \cr
\tau_{21}-b^2T^{-1}_{n-1}a^1 & \tau_{11}-b^2T^{-1}_{n-1}a^2 \cr}%
=D_{n-1}D_{n+1}%
\pmatrix{ T_{n-1} & a^1 & a^2 \cr
                               b^1  & \tau_{11} & \tau_{12} \cr
                               b^2  & \tau_{21} & \tau_{22} \cr} 
\]%
We will treated the last equality as the first Jacobi identity. By the same
technique it is not difficult to show that the following equality takes
place 
\[
D_{n}\pmatrix{ T_{n-1} & a^1 \cr
                  b^1 & \tau \cr}D_{n+1}%
\pmatrix{ T_{n-1} & a^1 & a^2 \cr
                                            d^1  & \nu & \mu \cr
                                           b^2  & \rho & \tau \cr}-D_{n}%
\pmatrix{ T_{n-1} & a^1 \cr
                  b^2 & \rho \cr}D_{n+1}%
\pmatrix{ T_{n-1} & a^1 & a^2 \cr
                                                       d^1  & \nu & \mu \cr
                                                     b^1 & \tau & \sigma \cr}%
= 
\]%
\[
D_{n}\pmatrix{ T_{n-1} & a^1 \cr
                  d^1 & \nu \cr}D_{n+1}%
\pmatrix{ T_{n-1} & a^1 & a^2 \cr
                                                       b^2  & \rho & \tau \cr
                                                     b^1 & \tau & \sigma \cr}
\]%
This equality we will use many times in what follows and will call second
Jacobi identity. These identities can be generalized on the case of
arbitrary semi-simple group. Reader can find these results in \cite{3}.

\subsection{Concrete calculations}

Let us take initial solution in the form 
\begin{equation}
Q=A=P=0,\quad B\equiv B(2),\quad E\equiv E(1).\quad D_{3}=-BE  \label{IS}
\end{equation}%
Application to this solution each of inverse transformations $T_{i}^{-1}$ is
mean less via arising zeroes in denominators. The chain of equations under
such boundary condition we will call as the chain with the fixed end from
the left *from one side).

The result of application to such initial solution $l_{3}$ times $T_{3}$
transformation looks as (for the checking of this fact only two Jacobi
identities of the previous subsection are necessary) 
\[
Q^{(l_{3}}=(-1)^{l_{3}-1}{\frac{\Delta _{l_{3}-1}}{\Delta _{l_{3}}}},\quad
D^{(l_{3}}=(-1)^{l_{3}}{\frac{\Delta _{l_{3}+1}}{\Delta _{l_{3}}}},\quad
\Delta _{0}=1 
\]%
\begin{equation}
A^{(l_{3}}=(-1)^{l_{3}}{\frac{\Delta _{l_{3}}^{B}}{\Delta _{l_{3}}}},\quad
P^{(l_{3}}={\frac{\Delta _{l_{3}}^{E}}{\Delta _{l_{3}}}},\quad \Delta
_{0}^{B}=\Delta _{0}^{E}=0  \label{T_3}
\end{equation}%
\[
B^{(l_{3}}={\frac{\Delta _{l_{3}+1}^{B}}{\Delta _{l_{3}}}},\quad
E^{(l_{3}}=(-1)^{l_{3}}{\frac{\Delta _{l_{3}+1}^{E}}{\Delta _{l_{3}}}},\quad
\Delta _{-1}=0. 
\]%
where $\Delta _{n}$ are minors of the n-th order of infinite dimensional
matrix 
\begin{equation}
\Delta =\pmatrix{ D & D_2 & D_{22} & .....\cr D_1 & D_{12} & D_{122} &
.....\cr D_{11} & D_{112} & D_{1122} & .....\cr ...... & ....... &
.........& .....\cr}  \label{DM}
\end{equation}%
and $\Delta _{l_{3}}^{E},\Delta _{l_{3}}^{B}$ are the minors of $l_{3}$
order in the matrices of which the last column (or row) is exchanged on the
derivatives of the corresponding order on argument $1$ of $E$ function (on
argument $2$ of the $B$ function in the second case).

In what follows the following notations will be used. $W^{l_{3},l_{1}}$, ($%
W^{l_{3},l_{2}}$)- the result of application discrete transformation $%
T^{l_{3}}T^{l_{1}}$ ($T^{l_{3}}T^{l_{2}}$) to the corresponding component of
the 3- wave field. $\Delta ^{l_{3},l_{1}}$ ($\Delta ^{l_{3},l_{2}}$) -
determinant of $l_{3}+l_{1}$ ($l_{3}+l_{2}$) orders, with the following
structure of its determinant matrix. The first $l_{3}$ rows (columns) of it
coincides with matrix of (\ref{DM}) and last $l_{1}$, $(l_{2})$ rows
(columns) constructed from the derivatives of $B$, ($E$) functions with
respect arguments 2, (1).

The result of additional application of $l_{1}$ times $T_{1}$ transformation
to the solution (\ref{T_3}) looks as 
\[
P^{(l_{3},l_{1}}={\frac{\Delta _{l_{3},l_{1}-1}}{\Delta _{l_{3},l_{1}}}}%
,\quad B^{(l_{3},l_{1}}={\frac{\Delta _{l_{3},l_{1}+1}}{\Delta
_{l_{3},l_{1}}}},\quad \Delta _{0}=1,\quad \Delta ^{l_{3},-1}\equiv \Delta
_{l_{3}}^{E} 
\]%
\begin{equation}
Q^{(l_{3},l_{1}}=(-1)^{l_{3}+l_{1}-1}{\frac{\Delta _{l_{3}-1,l_{1}}}{%
\Delta _{l_{3},l_{1}}}},\quad D^{(l_{3},l_{1}}=(-1)^{l_{3}+l_{1}}{\frac{%
\Delta _{l_{3}+1,l_{1}}}{\Delta _{l_{3},l_{1}}}},  \label{TT}
\end{equation}%
\[
E^{(l_{3},l_{1}}=(-1)^{l_{3}+l_{1}}{\frac{\Delta _{l_{3}+1,l_{1}-1}}{%
\Delta _{l_{3},l_{1}}}},\quad A^{l_{3},l_{1}}=(-1)^{(}{l_{3}+l_{1}}{\frac{%
\Delta _{l_{3}-1,l_{1}+1}}{\Delta _{l_{3},l_{1}}}}, 
\]%
We do not present the explicit form for components $W^{(}{l_{3},l_{2}}$,
which can be obtained without any difficulties from (\ref{TT}) by
corresponding exchanging of the arguments and unknown functions.

\section{Many-soliton solution of the scalar 3-waves problem}

The system (\ref{MS}) allows the following reducing ( under additional
assumption that all operators of differentiation are the real ones $\partial
_{\alpha }=\partial _{\alpha }^{\ast }$) 
\begin{equation}
P=B^{\ast },\quad A=E^{\ast },\quad Q=D^{\ast }  \label{REA}
\end{equation}%
In this case the system (\ref{MS}) is reduced to three equations 
\begin{equation}
B_{1}=-DE^{\ast },\quad E_{2}=-DB^{\ast },\quad D_{3}=-BE  \label{MSC}
\end{equation}%
for three complex valued unknown functions $(E,B,D)$.

Now we would like to demonstrate how the multi-soliton solutions of the
system (\ref{MSC}) may be obtained with the help of the technique of
discrete transformation in the most straightforward way.

With this aim let us consider the action of the direct and inverse $%
T_{i},T_{i}^{-1}$ transformations on the reduced solution of the system (\ref%
{MSC}). The trick consists in the fact that discrete transformation does not
conserve the condition of the reality (\ref{REA}) and starting from the
solution of the reduced system we come back to solution of irreductible one
and in some cases vice versa. We will denote the three dimensional vector $%
(Q,P,A)$ by the single symbol $\vec{Q}$ and the by symbol $\vec{D}$ three
dimensional vector $(D,B,E)$. Then the result of actions of direct and
inverse transformations on solution satisfying the condition of reality $%
\vec{Q}=\vec{D^{\ast }}$ is the following 
\[
T_{i}^{n}(\vec{D},\vec{D^{\ast }})=(t_{i})^{n}(\vec{q},\vec{d}),\quad
T_{i}^{-n}(\vec{D},\vec{D^{\ast }})=(\vec{d^{\ast }},\vec{q^{\ast }}) 
\]%
where $t_{i}$ are point like symmetries of the system (\ref{MS}) 
\[
t_{3}(Q,P,A,D,B,E)=(Q,-P,-A,D,-B,-E), 
\]%
\[
t_{2}(Q,P,A,D,B,E)=(-Q,-P,A,-D,-B,E), 
\]%
\[
t_{3}(Q,P,A,D,B,E)=(-Q,P,-A,-D,B,-E) 
\]%
It is obvious that $t_{i}^{2}=1$. Thus if we apply $2n$ times discrete
transformation to initial bad (nonreduced) solution $(0,\vec{D})$ and as a
result obtain $(t^{n}\newline
vec{D^{\ast }},0)$ then in the middle of the chain we will have solution
satisfying the condition of reality, which coincides with $n$ soliton
solution of the reduced system (\ref{MSC}).

The solution of the chain with the boundary conditions $\vec{Q}=0$ on the
left end of the chain and $\vec{D}=0$ on the right side we will call as the
chain with fixed ends. Really condition $\vec{D}=0$ is the system of
equation from which initial functions $D,B,E$ (see (\ref{MSC}) may be
defined as the solutions of ordinary differential equations (see Appendix
II).

\section{Matrix three waves problem in the space of three dimensions and its
discrete transformation}

In all calculations above we have never used (except of concrete resolving
of discrete transformation chains) the condition that operators of
differentiation are connected by the condition 
\[
\partial _{1}+\partial _{2}+\partial _{3}=0 
\]%
as it follows from the definition of this operators. So we can consider the
system (\ref{MS}) where all three operators are independent from each other
and correspond to differentiation with respect to one of coordinates of
three dimensional space. The second generalization consists in possibility
to consider the unknown function in (\ref{MS}) as the operator valued ones.
Of course in this case the order of the multiplications are essential and
exactly coincides with fixed by the formula (\ref{MS}).

$T_{3}$ discrete transformation in this case looks as 
\[
\bar{Q}=D^{-1},\quad \bar{A}=-BD^{-1},\quad \bar{P}=D^{-1}E, 
\]%
\[
\bar{B}=-D(BD^{-1})_{2},\quad \bar{E}=-D(D^{-1}E)_{1},\quad D^{-1}\bar{D}%
=QD-(D^{-1}D_{2})_{1} 
\]%
By the same technique for $T_{1}$ we have 
\[
\bar{P}=B^{-1},\quad \bar{Q}=B^{-1}A,\quad \bar{E}=-DB^{-1}, 
\]%
\[
\bar{D}=(DB^{-1})_{2}B,\quad \bar{A}=-B(B^{-1}A)_{3},\quad \bar{B}%
B^{-1}=BP-(B_{3}B^{-1})_{2} 
\]%
And at last the discrete transformation $T_{2}$ looks as 
\[
\bar{A}=E^{-1},\quad \bar{B}=E^{-1}D,\quad \bar{Q}=-PE^{-1}, 
\]%
\[
\bar{D}=-E(E^{-1}D)_{1},\quad \bar{P}=(PE^{-1})_{3}E,\quad E^{-1}\bar{E}%
=AE-(E^{-1}E_{3})_{1} 
\]%
As in the scalar case the discrete transformations in the case under
consideration are mutually commutative. The arising chains of equations for $%
(E,B,D)$ operator a valued functions ( the matrices of the finite dimensions
for instance) coincides with the investigated before matrix Toda chain.
Explicit solutions for this chains of equations with the fixed ends reader
can find in \cite{LY}. Uniting these results it is possible to construct
multi soliton solutions of the matrix 3-wave problem in three dimensions
similar to way proposed in \cite{LYY} for construction of multi soliton
solutions for matrix Devay-Stewartson equation.

\section{Outlook}

The concrete results of the present paper are concentrated in explicit
formulae for discrete transformations for 3-wave problem of the section two
and their generalization on the matrix case (section 6).

But a no less important is the understanding how the method of the discrete
transformation may be generalized on the case of multicomponent systems,
connected with the semi simple algebras of the higher ranks $r$. From
results of the present paper it is clear that in the case of arbitrary semi
simple algebra there are $r$ independent basis mutually commutative discrete
transformations. In what connection are this commutative objects with the
main ingredients of the representation theory of the group is very
interesting and intrigued question for further investigation.

And the last comment. The chain with two fixed ends can not be considered as
the basis for some finite dimensional representation of the group of the
discrete transformation, if it is at all possible to apply term group for it
in this case. On the function at the end point of chain it is impossible to
act by direct transformation at the right side and inverse on the left end.
What is discrete transformation from the group theoretical point of view in
this case? We at this time have no answer on this question.

\section{Acknowledgements}

The authors are indebted to CONNECUT for financial support.

\section{Appendix I}

In this appendix we would like to show, how it is possible to construct two-
dimensional integrable systems connected with $A_2$ algebra.

Let us consider the following $3\times 3$ polynomial matrix 
\begin{equation}
P(\lambda )=\pmatrix{ \tilde P^{11}_{n_1+1} & P^{12}_{n_2} & P^{13}_{n_3}
\cr P^{21}_{n_1} & \tilde P^{22}_{n_2+1} & P^{23}_{n_3} \cr P^{31}_{n_1} &
P^{32}_{n_2} & \tilde P^{33}_{n_3+1} \cr}  \label{DT}
\end{equation}%
where $P_{k}^{ij}$ are the polynomials of the degree $k$ (with respect to
parameter $\lambda $ and sign $\tilde{P}$ means that coefficient function on
the highest degree of the corresponding polynomial equal to unity.

Let us define coefficient function of all polynomials are defined from the
condition that between the columns of the matrix 
\[
\bar{P}=P(\lambda )\exp(\tau _{1}h_{1}+\tau _{2}h_{2}) 
\]%
takes place the linear dependence in $(n_{1}+n_{2}+n_{3}+3)$ points of the $%
\lambda $ plane, $h_{1},h_{2}$ Cartan elements of $A_{2}$ algebra and $\tau
_{i}=\phi _{i}(t,\lambda )+f_{i}(x,\lambda )$. $\phi _{i},f_{i}$ are
arbitrary rational functions with respect to argument $\lambda $.

The last condition is equivalent to the following system of linear equations
for defining the coefficient function (we present it here for elements of
the first row) 
\begin{equation}
\tilde{P}_{n_{1}+1}^{11}(\lambda _{s})+c_{s}\exp (\tau _{2}^{s}-2\tau
_{1}^{s})P_{n_{2}}^{12}(\lambda _{s})+d_{s}\exp -(\tau _{2}^{s}+\tau
_{1}^{s})P_{n_{3}}^{13}(\lambda _{s})=0  \label{CF}
\end{equation}%
\[
s=1,2,...,(n_{1}+n_{2}+n_{3}+3),\tau _{i}^{s}\equiv \tau (\lambda _{s}) 
\]%
$(n_{1}+n_{2}+n_{3}+3)$ exactly the number of coefficient function of
polynomials of the first row. Thus (\ref{CF}) is the linear system of
equation for their determination.

Let us now determinant $Det(P(\lambda )=Det(\bar{P}(\lambda ))$. From (\ref%
{DT}) it follows that it is the polynomial of $(n_{1}+n_{2}+n_{3}+3)$ degree
with unity coefficient before the highest term and from condition (\ref{CF})
that it has zeroes in $(n_{1}+n_{2}+n_{3}+3)$ points $\lambda _{s}$ of the $%
\lambda $ plane. Thus 
\begin{equation}
Det(P(\lambda ))=Det(\bar{P}(\lambda ))=\Pi
_{k=1}^{(n_{1}+n_{2}+n_{3}+3)}(\lambda -\lambda _{k})  \label{DD}
\end{equation}%
Now let us calculate the matrix $\dot{\bar{P}}{\bar{P}}^{-1}$, where $\dot{f}
$ means the differentiation with respect to one of two independent arguments
of the problem $x,t$. From the definition of the inverse matrix it follows
that matrix elements of this matrix are the following ones 
\begin{equation}
(\dot{\bar{P}}{\bar{P}}^{-1})_{\alpha ,\beta }={\frac{Det(P_{\beta
}\rightarrow \dot{P}_{\alpha }+P_{\alpha }(\dot{\tau}_{i+1}-\dot{\tau}_{i})}{%
\Pi _{k=1}^{(n_{1}+n_{2}+n_{3}+3)}(\lambda -\lambda _{k})}},\quad \tau
_{0}=\tau _{3}=0  \label{ME}
\end{equation}%
This symbolical form means that the determinant matrix of numerator arises
after exchanging of the $\beta $ row of the $P$ matrix on the $\alpha $ row
of the matrix $\dot{\bar{P}}\exp -(\tau _{1}h_{1}+\tau _{2}h_{2})$.

It is not difficult to understand that matrix $\dot{\bar{P}}{\bar{P}}^{-1}$
possess all the same singularities as functions $\tau $ by themselves.

Now let us illustrate situation on the example of three wave interaction,
choosing $\tau _{1}=\lambda (c_{1}t+c_{2}x),\tau _{2}=\lambda
(d_{1}t+d_{2}x) $. Let us calculate in this case for instance $(\bar{P}_{t}{%
\bar{P}}^{-1})_{11}$. In connection with (\ref{ME}) we numerator determinant
have 
\[
Det%
\pmatrix{
\dot {\tilde P_{n_1+1}}+\tilde P_{n_1+1}c_1\lambda & \dot {P_{n_2}}+P_{n_2}
(c_2-c_1)\lambda & \dot {P_{n_3}}- P_{n_3}c_2\lambda \cr
P_{n_1} & \tilde P_{n_2+1} & P_{n_3} \cr
P_{n_1} & P_{n_2} & \tilde P_{n_3+1} \cr} 
\]%
It is obvious that between the columns of the matrix $\dot{\bar{P}}$ the
linear dependence takes place with the same coefficients and so numerator
determinant has zeroes in the same points as determinant in enumerator.
Computation the degrees of numerator shows that it is polynomial of the $%
(n_{1}+n_{2}+n_{3}+4)$ order and so considered matrix element is the linear
function of the $\lambda $ parameter. From (\ref{DD}) it follows that it
equal exactly $c_{1}\lambda $. The same not conversion calculations show
that matrix $\bar{P}_{t}{\bar{P}}^{-1}$ coincides with the $u$ matrix from
the introduction after identification 
\[
P=(P^{12})_{n_{2}}^{n_{2}},\quad Q=(P^{13})_{n_{3}}^{n_{3}},\quad
B=(P^{21})_{n_{1}}^{n_{1}}, 
\]%
\[
A=(P^{23})_{n_{3}}^{n_{3}},\quad D=(P^{31})_{n_{1}}^{n_{1}},\quad
E=(P^{32})_{n_{2}}^{n_{2}} 
\]%
where values above are coefficients at the highest degree terms of the
corresponding polynomial. These terms are known from the solution of the
linear system (\ref{CF}) and so we have explicit solution of the system (\ref%
{3W}).

\section{Appendix II}

In this appendix we would like to consider the simple example of soliton
solution of 3-wave problem. We specially consider this simplest example in
details to give the reader possibility to feel self-consistent of the whole
construction of the present paper.

Let in notations of the 5-th section $l_{3}=2,l_{1}=0$. Condition that
vector $\vec{D^{2,0}}=0$ is equivalent to the following system of equations 
\begin{equation}
\Delta _{3}=\Delta _{3}^{B}=\Delta _{3}^{E}=0  \label{SE}
\end{equation}%
The first of this equations leads uniquely to explicit form of initial $D$
function 
\begin{equation}
D=\phi _{1}(1)f_{1}(2)+\phi _{2}(1)f_{2}(2),\quad \phi _{1}=\phi ^{\prime
},\quad f_{2}=\dot{f}  \label{DDD}
\end{equation}%
Using the initial conditions (\ref{IS}) equation $\Delta _{3}^{E}=0$ may be
rewritten consequently 
\[
B\Delta _{3}^{E}=-Det%
\pmatrix{D & D_2 & D_1 \cr
                                                D_1 & D_{12} & D_{11} \cr
                                             D_{11} & D_{112} & D_{111} \cr}= 
\]%
\[
(\dot{f}_{1}f_{2}-\dot{f}_{2}f_{1}) Det%
\pmatrix{ \phi_1 & \phi_2 & \phi'_1f_1+\phi'_2f_2 \cr
                     \phi'_1 & \phi'_2 & \phi''_1f_1+\phi''_2f_2 \cr
                     \phi''_1 & \phi''_2 & \phi'''_1f_1+\phi'''_2f_2 \cr} 
\]%
Keeping in mind that $\phi ,f$ are the functions of the different arguments
we conclude the last equations is equivalent to equality to zero of the two
determinants of third order. The last condition in its turn can be rewritten
as the system of equations 
\[
\phi _{1}^{\prime }=p\phi _{1}+q\phi _{2},\quad \phi _{2}^{\prime }=s\phi
_{1}+t\phi _{2} 
\]%
\begin{equation}
\phi _{1}^{\prime \prime }=p\phi _{1}^{\prime }+q\phi _{2}^{\prime },\quad
\phi _{2}^{\prime \prime }=s\phi _{1}^{\prime }+t\phi _{2}^{\prime }
\label{PQRS}
\end{equation}%
\[
\phi _{1}^{\prime \prime \prime }=p\phi _{1}^{\prime \prime }+q\phi
_{2}^{\prime \prime },\quad \phi _{2}^{\prime \prime \prime }=s\phi
_{1}^{\prime \prime }+t\phi _{2}^{\prime \prime } 
\]%
From (\ref{PQRS}) it follows immediately that ($\phi _{2}\neq c\phi _{1}$) $%
p^{\prime }=q^{\prime }=s^{\prime }=t^{\prime }=0$ and functions $\phi
_{1,2} $ are the solutions of the first row of (\ref{PQRS})-the linear
system of equation with the constant coefficients. Solution of this system
is obvious 
\[
\phi _{1}=c_{1}\exp \lambda _{11}+c_{2}\exp \lambda _{21},\quad \phi
_{2}=c_{3}\exp \lambda _{11}+c_{4}\exp \lambda _{21}.
\]%
From the equation $E\Delta _{3}^{B}$ by the same way we obtain 
\[
f_{1}=d_{1}\exp \mu _{12}+d_{2}\exp \mu _{22},\quad f_{2}=d_{3}\exp \mu
_{12}+d_{4}\exp \mu _{22}, 
\]%
where $c,d,\lambda ,\mu $ arbitrary numerical parameters.

The initial conditions 
\[
-D_{3}=D_{1}+D_{2}=BE\equiv (b_{1}\exp \mu _{1}2+b_{2}\exp \mu
_{2}2)(e_{1}\exp \lambda _{1}1+e_{2}\exp \lambda _{2}1) 
\]%
allow using (\ref{DDD}) allow determine parameters $b,e$ and find one
relation connected parameters $c,d,\lambda ,\mu $. Now let us calculate
vector $Q^{2,0}$ using explicit expressions for $D,B,E$ functions. The last
two ones we present in the following form $E=p\phi _{1}+q\phi
_{2},B=rf_{1}+sf_{2}$. 
\[
Q^{2,0}=-{\frac{D}{D_{2}}}=-{\frac{\phi _{1}f_{1}+\phi _{2}f_{2}}{D%
\pmatrix{\phi_1 & \phi_2 \cr
                 \phi'_1 & \phi'_2 \cr}D%
\pmatrix{f_1 & \dot f_1 \cr
                                                                                 f_2 & \dot f_2 \cr}}} 
\]%
\[
P^{2,0}={\frac{D\pmatrix{f_1 & p \cr
                                    f_2 & q \cr}}{D%
\pmatrix{f_1 & \dot f_1 \cr
                                                                                       f_2 & \dot f_2 \cr}}},\quad A^{2,0}={\frac{D%
\pmatrix{\phi_1 & \phi_2 \cr
                                r & s \cr}}{D%
\pmatrix{\phi_1 & \phi_2 \cr
                                                          \phi'_1 & \phi'_2 \cr}}} 
\]%
Conditions of reality leads to other restriction on parameters involved. It
is clear that two possibilities in the choice of parameters $\lambda $ and 
$\mu$ are possible $\lambda _{2}=-\lambda _{1}^{\ast }$, $\lambda _{1}=-\lambda
_{1}\ast$ and $\lambda _{2}=-\lambda _{2}\ast $. And the same limitations on
parameters $\mu _{2}$. We do not present here explicit form for the other
restrictions. This is pure algebraic manipulations.

\end{document}